\documentclass[aps,showkeys,showpacs,notitlepage,superscriptaddress]{revtex4-1}
\pdfoutput=1
\usepackage{mathtools}
\usepackage{amsmath}
\usepackage{amssymb}
\usepackage{graphicx}
\usepackage{hyperref}
\usepackage{color}
\usepackage[utf8]{inputenc}

\begin{document}

\title{Network approach towards understanding the crazing in glassy amorphous polymers}
\author{Sudarkodi Venkatesan}
\thanks{SV and RPV contributed equally to this work}
\affiliation{The Institute of Mathematical Sciences, Homi Bhabha National Institute, Chennai, India}
\affiliation{Department of Mechanical Engineering, Indian Institute of Technology, Kanpur, India}
\author{R.P. Vivek-Ananth}
\thanks{SV and RPV contributed equally to this work}
\affiliation{The Institute of Mathematical Sciences, Homi Bhabha National Institute, Chennai, India}
\author{R.P. Sreejith}
\affiliation{The Institute of Mathematical Sciences, Homi Bhabha National Institute, Chennai, India}
\author{P. Mangalapandi}
\affiliation{The Institute of Mathematical Sciences, Homi Bhabha National Institute, Chennai, India}
\author{Ali A. Hassanali}
\email{Corresponding author: ahassana@ictp.it}
\affiliation{The Abdus Salam International Centre for Theoretical Physics, Trieste, Italy}
\author{Areejit Samal}
\email{Corresponding author: asamal@imsc.res.in}
\affiliation{The Institute of Mathematical Sciences, Homi Bhabha National Institute, Chennai, India}

\begin{abstract}
We have used molecular dynamics to simulate an amorphous glassy polymer with long chains to study deformation mechanism of crazing and associated void statistics. The Van der Waals interactions and the entanglements between chains constituting the polymer play a crucial role in crazing. Thus, we have reconstructed two underlying weighted networks, namely, the Van der Waals network and the Entanglement network from polymer configurations extracted from the molecular dynamics simulation. Subsequently, we have performed graph-theoretic analysis of the two reconstructed networks to reveal the role played by them in crazing of polymers. Our analysis captured various stages of crazing through specific trends in the network measures for Van der Waals networks and entanglement networks. To further corroborate the effectiveness of network analysis in unraveling the underlying physics of crazing in polymers, we have contrasted the trends in network measures for Van der Waals networks and entanglement networks in the light of stress-strain behaviour and voids statistics during deformation. We find that Van der Waals network plays a crucial role in craze initiation and growth. Although, the entanglement network was found to maintain its structure during craze initiation stage, it was found to progressively weaken and undergo dynamic changes during the hardening and failure stages of crazing phenomena. Our work demonstrates the utility of network theory in quantifying the underlying physics of polymer crazing and widens the scope of applications of network science to characterization of deformation mechanisms in diverse polymers.
\end{abstract}

\pacs{89.75.Fb; 83.10.Rs; 81.05.Lg; 62.20.Fe}
\maketitle

\section{Introduction}
\label{introduction}

Polymers have become an indispensable component of our everyday life \cite{Chen1993,Kenawy2007,Clough2001,Kamcev2016}. They have made possible new generation of materials which are used in automotive, construction, and medical applications among others \cite{Bakis2002,Mortazavian2015,Maitz2015}. The commercial importance of the lifetime of polymers under usage has led to extensive studies on deformation characteristics and such insights can impact efforts to design new polymers \cite{Kinloch1983,Qian2000,Hirai2004}. The wide range of applications are a result of the diverse morphologies and properties exhibited by polymers, from soft gels to hard plastics \cite{Wichterle1960,Chujo1992,Tarascon1996,Erbil2003}. The diverse properties exhibited by polymers are a direct outcome of their inherent structure and energetics. At the level of structure, polymers can be crosslinked or uncrosslinked, linear or branched, amorphous or crystalline, which determine their properties. At the level of energetics, the bonded and non-bonded interactions between the atoms determine the conformation, flexibility and dynamics of the polymer, which in turn reflect in the overall behavior of the polymer. Hence, a major goal of material science is to better understand the link between the polymer structure and energetics, and that of the properties exhibited by them \cite{Jancar2010,Hu2012,Appel2014}. Towards this goal, molecular dynamics (MD) simulation serve as an invaluable tool to probe the polymer behaviour at an atomic scale (\cite{Barrat2010,Li2013,Li2014}). Particularly, MD simulations have been widely used to study the deformation of polymers \cite{Kenward2004,Venkatesan2015,Xin2015}. Subsequent analysis of samples from MD simulation at different stages of deformation have revealed new insights on polymer failure mechanism \cite{Stevens2001,Gersappe2002,Chenoweth2005,Venkatesan2015}. Advances in the computing power at disposal and hence the possibility of studying ever larger systems has led to a deluge of data. A typical MD simulation of polymer deformation considers systems in the order of $10^5-10^7$ atoms (\cite{Vu2014}). The analysis of the resulting trajectories of the atoms of the polymer sample is essential to understand the micromechanical aspects of the deformation process. Studying micromechanical aspects which largely depend on the response of interconnected structure of polymers due to imposed deformation conditions warrants an appropriate tool. Such an effective tool to mine data from MD simulation of microscale configurations is thus vital to develop rigorous multiscale models. In this work, we employ network theory \cite{Watts1998,Barabasi1999,Albert2002,Barabasi2004,Samal2006,Newman2010,Fortunato2010,Dorogovtsev2013} as a tool to analyze samples from MD simulation of polymer deformation to reveal new insights on failure mechanism. Previously, network theory based approaches have been utilized to analyze the vast data generated from MD simulations of proteins \cite{Ghosh2007, Sethi2009, Vanwart2012, Ribeiro2015}.

Network science \cite{Watts1998,Barabasi1999,Albert2002,Barabasi2004,Samal2006,Newman2010,Fortunato2010,Dorogovtsev2013} has made rapid advances in diverse fields. In the last two decades, network theory has been extensively used to analyze diverse social, biological and technological networks, and these studies have been successful in unraveling the organization of diverse complex networks
\cite{Watts1998,Barabasi1999,Albert2002,Barabasi2004,Samal2006,Newman2010,Fortunato2010,Dorogovtsev2013}. Network analysis is a promising framework to investigate the structure and dynamics of underlying complex networks associated with polymer samples from MD simulation of deformation which are expected to play a key role in their failure. However, very few studies have employed network theory to reconstruct and analyze underlying networks associated with polymer samples from MD simulations of polymer transitions \cite{Bohbot2004,Billen2009,Gavrilov2014}. For example, Bohbot {\it et al} \cite{Bohbot2004} have employed network measures to analyze pre-gel association of telechelic molecules. Billen {\it et al} \cite{Billen2009} have used network theory to reveal new insights on the transition from sol to gel structure of telechelic polymer solutions. Billen {\it et al} \cite{Billen2009} represented the telechelic polymer solution as a network where the cross-linkages between chains are taken as nodes while chains connecting cross-linkages are taken as edges. Billen {\it et al} \cite{Billen2009} studied the degree distribution of the underlying networks associated with telechelic polymer solution as a function of the temperature. Moreover, Billen {\it et al} \cite{Billen2009} showed that the transition from sol to gel is reflected by change in degree distribution of the underlying networks associated with telechelic polymer from bimodal to unimodal with high degree. Gavrilov {\it et al} \cite{Gavrilov2014} have analyzed a representation of a crosslinked epoxy molecular simulation sample, with the molecules as nodes and the crosslinks as edges. In Gavrilov {\it et al} \cite{Gavrilov2014}, the features of the epoxy network such as multiple edges, simple cycles and complex cycles were analyzed, and their distribution was studied as a function of the sample sizes used in MD simulation. We here employ network theory to study crazing in polymers, and to the best of our knowledge, this is the first attempt to use network science to understand the underlying physics of crazing in polymers. Mesoscale models of polymer deformation are based on two vital interactions underlying polymer dynamics, namely Van der Waals interactions and the entanglement between chains \cite{Termonia1987,Reddy2008}. Such models have successfully captured key features of polymer deformation like yielding and hardening. This motivates our choice of underlying networks to reconstruct and analyze from MD simulations of polymer deformation.

In this study we analyze an amorphous glassy polymer under deformation, exhibiting crazing \cite{Kambour1964,Kramer1983}. Crazing is typically exhibited by polymers with low entanglement density, where the deformation zone evolves into a fibrilated network as a precursor to crack growth \cite{Doll1990}. We study the crazing in a model polyethylene sample \cite{Fukuda1997} using MD simulations. The chain segments of the polymer interact with each another through Van der Waals interactions, and it has been shown that both the entanglements and the Van der Waals interactions play a crucial role in the dynamics of polymer deformation \cite{Kramer1983,Kardomateas1985,Doll1990,Rottler2003,Deblieck2011,Venkatesan2015}. In this work, we perform a network level analysis of the Van der Waals network and the Entanglement network associated with polymer samples from MD simulation that were extracted at different stages of crazing phenomena. By investigating a wide range of network measures in Van der Waals networks and entanglement networks, we identify network-based signatures that capture important characteristics of the crazing phenomena. Our detailed analysis of Van der Waals networks and entanglement networks demonstrate the effectiveness of network theory in unraveling the dynamics of polymer deformation.

The remaining paper is organized as follows. In section \ref{methods}, we describe methods associated with polymer sample preparation, MD simulation of amorphous model polymer, estimation of void structure in polymers, and identification of entanglements between chains of a polymer. In section \ref{obs-md}, we summarize observations from MD simulation of the crazing phenomena and relate them to evolution of void structure in polymers. In section \ref{reconstruction}, we describe our method to reconstruct two underlying networks associated with polymers: Van der Waals network and entanglement network. In section \ref{networkanalysis}, we present results from graph-theoretic analysis of Van der Waals networks and entanglement networks, and relate them to the underlying physics of crazing in polymers. In section \ref{summary}, we conclude by summarizing our results and providing suggestions for future work.


\section{Molecular dynamics simulation and associated methods}
\label{methods}

Molecular dynamics (MD) simulation typically model the dynamics of a polymer at the atomic scale. In such simulations, the energetics are governed by carefully designed forcefields which mimic the bonded and non-bonded interactions between atoms of a polymer. Subsequently, the results of the MD simulation are analyzed to draw conclusions on the properties of the polymer. In the following sections, we describe the polymer sample preparation and deformation procedure. Moreover, the growth and coalescence of voids is a major deformation characteristic exhibited during polymer crazing. We describe below the method used to determine voids in the polymer sample. Moreover, we are also interested in the network of entanglements in the considered polymer, and thus, we describe below the method used to identify entanglements in the polymer sample. Further to better understand the role of entanglements in dynamics of polymer deformation, we track at different stages the changes in the network of entanglements within the sample.

\subsection{Initial sample preparation}

As our focus is crazing phenomenon in polymers, we consider a amorphous glassy polymer which crazes upon deformation. We followed the procedure described in previous work \cite{Venkatesan2015} to prepare amorphous samples of polyethylene. Though polyethylene exhibits semi-crystalline structure by nature, an amorphous sample was prepared as a model system to study polymer crazing. We consider the -$CH_2$- entity in the polyethylene as an united atom. Thus, every ethylene monomer is taken as two bonded united atoms. Our polymer sample has 160 chains with 1000 united atoms each, and this ensures that the prepared sample has sufficient entanglement density to exhibit crazing. To prepare the initial sample, individual chains of varying configurations were packed in random orientations in a rectangular box of dimensions 100$\AA$ $\times$ 100$\AA$ $\times$ 500$\AA$. Subsequently, the forcefields specified in Table \ref{ffield} and detailed in the next section were used for the MD simulation. The MD simulation of the polymer sample with periodic boundary conditions was carried out in LAMMPS software \cite{Plimpton1995}. Using the NVT ensemble, the initial polymer sample was equilibrated whereby the volume and temperature of the sample were kept constant. This equilibration step ensures removal of any close contacts between chains that may arise while preparing the initial sample. A temporary softer Lennard-Jones ($LJ$) potential was used during this equilibration step to slowly move the atoms in the initial sample to their equilibrium positions. Furthermore, double rebridging algorithm, a Monte carlo technique, was used to correct the configurations of the chains in the initial sample as described in Ref. \cite{Chen2013}, whereby the average configuration of the chains in the initial sample was made closer to the expected configuration of a polyethylene chain.

Apart from structural corrections, the initial sample needs to be at the desired temperature and pressure. Thus, NPT ensemble was used for further equilibration of 1-2 ns, where the initial sample was simulated at temperature equal to 100 K and pressure equal to 1 bar till the fluctuations in the temperature and pressure profiles were minimized. Thus, properly equilibrated atomic samples replicating the characteristic features of the polymer like density, characteristic ratio and pair distributions at required temperature and pressure were used for our deformation studies. Due to the amorphous nature of the considered polymer, any variations in the configuration of the initial sample used for MD simulation may lead to observed differences in the results of the deformation process. In order to understand the dependence of results on starting configuration of the MD simulation, we prepared 10 different initial polymer samples which were further equilibrated for deformation analysis. In this work, our results and conclusions are based on the MD simulations of these 10 different initial samples.
We remark that the results for the MD simulation of one initial sample is shown in main text while the corresponding results for remaining 9 initial samples are included in the supplementary information (SI).


\subsection{Molecular dynamics simulation}

In this study, we deform the polymer sample under an imposed triaxial stress state, which is essential for craze formation. For our simulations, we use LAMMPS software \cite{Plimpton1995} along with the forcefields specified in Table \ref{ffield}  which were taken from Fukuda and Kuwajima \cite{Fukuda1997}. It is noteworthy that Fukuda and Kuwajima \cite{Fukuda1997} have successfully used the forcefields in Table \ref{ffield} to predict the experimentally measured diffusion coefficients for methane in polyethylene. The bonds between the consecutive atoms $i$ and $j$ in a chain are governed by a harmonic potential $E_b$ where the equilibrium distance is given by $r_0$. The angle between three consecutive and connected atoms $i$, $j$ and $k$, $\theta_{[ijk]}$, is governed by the potential $E_\theta$. The bond length and angle between consecutive atoms are governed by strong potentials which ensure that the equilibrium values are maintained. However, the dihedral angle between four consecutive and connected atoms $i$, $j$, $k$ and $l$, $\phi_{[ijkl]}$, has a relatively weaker potential $E_\phi$ which permits shifts in the chain configuration. Apart from these bonded interactions, non-bonded interactions $U$, govern the energetics between non-bonded atoms within the same chain and other chains. This potential function $U$ can be attractive or repulsive depending on the distance between two atoms, and this force ultimately ensures an equilibrium distance between two atoms.

\begin{table}[!ht]
\caption{Specification of forcefields for the united atom model of polyethylene (\cite{Fukuda1997}) used in the molecular dynamics simulation.}
\centering
\scriptsize
\begin{tabular}{|p{2.0cm}|c|p{4cm}|}
\hline
Forcefield type & Equation & Parameters\\
\hline
\multicolumn{3}{|c|}{Stretching}\\
\hline
 Harmonic & \scriptsize $E_b(r_{[ij]}) = \frac{1}{2} k_b \left ( r_{[ij]} - r_0 \right )^2$ \; & \scriptsize $k_b=2745 \; \mathrm{kJ/mol \AA^2}$ and $r_0 = 1.53 \; \mathrm{\AA}$.\\

\hline
\multicolumn{3}{|c|}{Bending}\\
\hline
 Harmonic & \scriptsize $E_\theta (\theta_{[ijk]})= \frac{1}{2}k_\theta \left ((\theta_{[ijk]}) - (\theta_0)\right )^2$\; & \scriptsize $k_\theta=748.9 \; \mathrm{kJ/mol/rad^2}$ and $\theta_0 = 113.3^o$.\\
\hline
\multicolumn{3}{|c|}{Dihedral}\\
\hline
Cosine & \scriptsize $E_\phi(\phi_{[ijkl]}) = \frac{1}{2} \sum_{n=1}^{3} A_i \left ( \cos (n \phi_{[ijkl]}) \right )$\; & \scriptsize  $A_1 = 7.86$,  $A_2 = 4.36$ and \\
& &\scriptsize $A_3 = 15.56 \; \mathrm{kJ/mol}$ \\
\hline
\multicolumn{3}{|c|}{Pair interactions}\\
\hline
$CH_3-CH_2$  LJ & $U(r)=4\epsilon \; [\frac{\sigma^6}{r^6}-\frac{\sigma^{12}}{r^{12}}]$ & $\epsilon=0.88\;$kJ/mol,$\sigma=3.76 \; \AA $   \\
$CH_2-CH_2$  LJ  & $U(r)=4\epsilon \; [\frac{\sigma^6}{r^6}-\frac{\sigma^{12}}{r^{12}}]$ & $\epsilon=0.36\;$kJ/mol,$\sigma=4.06 \; \AA $   \\
\hline
\end{tabular}
\label{ffield}
\end{table}

The initial sample after equilibration was deformed at a constant velocity of 0.0003 $\AA\ fs^{-1}$ along the longer dimension of the sample. This ensures that during craze growth in polymer sample there is sufficient material above and below the craze to enable flow of chains into the crazing layer. During deformation process, the other sides of the polymer sample were kept fixed to produce a triaxial stress state which is essential for craze formation \cite{Venkatesan2015}. As mentioned earlier, we performed MD simulations starting from 10 different initial samples, and each sample was subjected to the same deformation rate. The simulation runs for 10 different initial samples were carried out till the stress drops sharply to below 80 MPa (indicating failure), and these runs typically lasted for 7-8 ns. Since the polymer sample is discrete in nature, estimating stress, a continuum property, requires custom-built measures that have been developed for discrete atomic samples. The virial stress is one such stress measure for discrete samples which has been extensively used due to its simplicity. A review of stress measures in Ref. \cite{Admal2010}, showed that the virial stress must be averaged over a significant time span to ensure proper stress estimation. Hence, we use averaged virial stress here to measure the stress in the polymer sample.

In Figure \ref{stress-strain-curve}, we show the evolution of the stress for one of the initial samples in the MD simulation at different strain values during polymer deformation. In SI Figure S1, we show the stress-strain curve from the MD simulations of all the 10 different initial samples considered here. To study the deformation characteristics of crazing in polymer, we have extracted 25 configurations across different stages of crazing phenomena from the MD simulation of each sample (Figure \ref{stress-strain-curve}(a)). Among the 25 selected configurations, the first 4 closely spaced configurations correspond to the initial yielding of polymer sample which occurs over a narrow strain interval, and these 4 configurations can be used to study the polymer deformation characteristics both before and after the yield point (Figure \ref{stress-strain-curve}(a)). The remaining 21 configurations were taken at equally spaced intervals till the sudden drop in stress to very low values signifying final failure of the polymer sample (Figure \ref{stress-strain-curve}(a)).


\subsection{Estimation of voids in the polymer sample}
\label{void-method}

The evolution of the void structure is an important characteristic of plastic deformation \cite{Fokoua2014,Gurson1977}. The evolution, growth and coalescence of microvoids plays a crucial role in the craze initiation process \cite{Argon2011,Ichinomiya2017}. Subsequently the craze structure evolves as a highly voided fibrillated structure as observed in the crazed polymer sample configurations in Figure \ref{stress-strain-curve}. The profile of the fractional volume occupied by the polymer was studied by Venkatesan and Basu \cite{Venkatesan2015} along the deformation direction at different strain values  whereby regions of marked drop in fractional volume were observed indicating voids. Hence, there is a considerable interest in understanding the evolution of the voids during growth and failure of the craze. The role of void formation in initial yielding and softening phases of crazing phenomena in polymers has been recently studied by Ichinomiya {\it et al} \cite{Ichinomiya2017}. However, the role of void structure in the growth and failure phases of crazing phenomena in polymers has not been investigated. Furthermore, the underlying void structure of the observed honeycomb craze structure in polyethylene \cite{Venkatesan2015} has also not been investigated. Thus, it is important to investigate the role of voids in the growth and failure phases of crazing phenomena in polymers. Moreover a proper understanding of the void structure will aid in future development of more realistic plastic deformation models for crazing in comparison to idealized models \cite{Haward1973}.

In order to identify the voids, the polymer sample was mapped onto a three dimensional mesh of 1$\AA$ dimension. Subsequently, the sample was discretized into cubes of dimension 1$\AA$ $\times$ 1$\AA$ $\times$ 1$\AA$ where the cuboidal discretization facilitates partitioning of the three dimensional rectangular shape. Moreover, the effective volume occupied by an united atom in our polyethylene sample is nearly 21$\AA^3$, and thus, the chosen discretization of 1$\AA^3$ is adequate to obtain sufficiently accurate results. The effective volume occupied by an united atom in the polymer sample is based on the equilibrium distance of the LJ potential between any pair of united atoms. At equilibrium distance between two atoms, the potential energy of interaction between them is minimum. When a system of atoms is densely packed and equilibrated, the atoms will occupy positions such that the distance between them is the equilibrium distance. Based on the coordinates of united atoms and the effective volume of an united atom of the polymer sample, we determine the regions of the mesh which are occupied by atoms. Afterwards the regions of the mesh which were not occupied by atoms were considered to be empty space. Note that we consider a cube of the mesh to be occupied even if it is partially filled with atoms. We next determine adjacent cubes of the mesh which are empty and group such empty cubes into a single void. In order to group the adjacent empty cubes of mesh into a single void, we have employed the density-based spatial clustering of applications with noise (DBSCAN) \cite{Ester1996} algorithm within the Python package scikit-learn \cite{Pedregosa2011}. We have determined the number of voids with their respective volumes in the 25 selected configurations of the polymer sample, and this dataset was used to investigate the evolution of void structure (void size distribution) at different stages of crazing phenomenon (Figure \ref{void-energy}; SI Figures S2-S13). Note that the attractive part of the potential between atoms of a polymer sample remains active for long distance, and thus, smaller voids are more likely to disappear during the course of deformation. A stable void is expected to be sufficiently large to ensure adequate separation between the surrounding atoms to render the attractive potential force between atoms on different sides of the void to be ineffective.

\subsection{Identification of entanglements}

\begin{figure}
\includegraphics[width=.63\columnwidth]{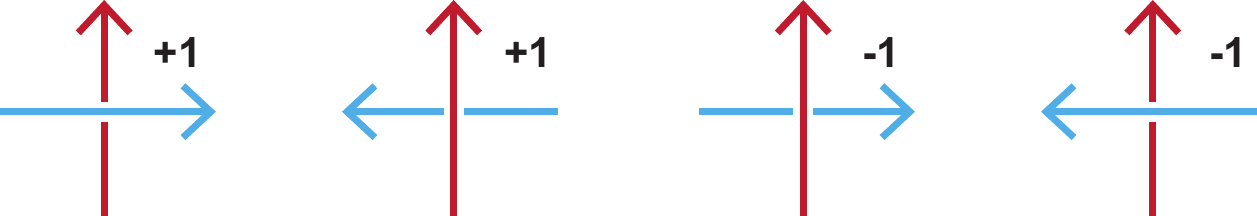}
\caption{A schematic of a criss cross point between two segments detailing the nature of the overlap and their orientation for estimating link number. The two chain segments are shown as arrows of red and blue color. The underlying segment is indicated by a broken line. The arrow head indicates the direction of the chain segment at a given criss cross point. In this study we take the chain direction along increasing monomer number. The method of assignment of -1/+1 at the criss cross point is shown. The summation of the assigned values at all criss cross points between the two segments gives the link number.}
\label{linknum}
\end{figure}

Entanglements are topological constraints between chains that play an important role in crazing. In previous work \cite{Venkatesan2015}, one of us has used the linking number method \cite{Adams2004} to identify entanglements between different chains in the polymer sample. The linking number methodology successfully identified entanglements in the polymer sample and the observed average entanglement length was found to match with the polyethylene entanglement length \cite{Venkatesan2015}. Briefly, the linking number method to identify entanglements in polymer sample can be described as follows. At first segments of a pair of chains that are closely placed, within interbead distance of 7$\AA$, in the polymer sample were identified. Next the closely placed chains were unfolded and the minimum images of a pair of closely placed chain segments was considered for further examination. Note that the linking number is a characteristic of knots. As a pair of chain segments examined here behave like tangles, their end points were hypothetically considered to be linked to form a knot. Since the pair of chain segments also criss cross each other, such points were noted by projecting the sample configurations on to a plane. At each criss cross point, a value of +1 or -1 was assigned depending on the nature and orientation of the overlap between chain segments as is schematically shown in Figure \ref{linknum}. In Figure \ref{linknum}, the blue and red lines indicate the two chain segments at a criss cross point, and the arrow heads indicate the direction of the chain segment, which here is taken to be along the increasing monomer number. The summation of these values for a pair of chain segments divided by two gives the linking number. Since the chain segments considered here are tangles, the summation over linking number may not always be a multiple of two. Moreover, a linking number of zero indicates that the segments are not linked and can move away without imposing any constraint on each other. However, a linking number greater than zero indicates entanglement and its extent increases with the value of the linking number. Subsequently, the estimated linking number for a pair of chain segments was used to verify whether the segments were entangled (i.e, if the linking number is greater than zero). Lastly, the closest criss cross point of contact between the two chain segments was taken as the monomer number where the entanglement occurs. As detailed in next section, we use this information on the monomer number where two chains entangle to track the dynamic changes in entanglements during polymer deformation. As we describe in section \ref{reconstruction}, entanglements are considered as nodes in the reconstruction of the entanglement network (Figure \ref{netrecon}).


\subsection{Dynamic changes in entanglements with polymer deformation}

Entanglements constrain the movement of chain segments and the extent of entanglements can be diverse in a polymer sample. Thus, some chain segments can be highly intertwined or entangled in a polymer sample and extricating such segments is very difficult, while other chain segments may be weakly intertwined or entangled and extricating such segments is much easier. Weak entanglements are highly sensitive to changes in the orientation of constituent segments upon polymer deformation. As the polymer sample deforms, the strong entanglements remain intact while segments constituting weak entanglements disentangle or undergo alteration due to changes in their orientation. In summary, new entanglements may appear while old entanglements may disappear during the deformation of the polymer sample. In order to study the birth and death of entanglements during the course of MD simulation, we use the information obtained from linking number method on (closest) monomer number where two segments entangle for each entanglement in the polymer sample. This information on
the pair of chains forming an entanglement and the monomer number of closest contact between chains can be used to distinguish between old entanglements and new entanglements while comparing configurations at different stages of polymer deformation.

Specifically, we have studied the appearance of new entanglements and disappearance of old entanglements between two consecutive configurations among the 25 selected configurations across different stages of crazing phenomena from the MD simulation of each of the 10 different initial samples (SI Figures S14-S23). We determine old and new entanglements between two consecutive configurations as follows. Firstly, we compare the identity of chains that form an entanglement in earlier configuration with the identity of chains that form an entanglement in later configuration. If the identity of chains forming the considered entanglement in earlier configuration matches with that forming the considered entanglement in later configuration, we compare the monomer positions of the considered entanglements in the two consecutive configurations. If the monomer positions of the considered entanglements are also same in the two consecutive configurations, the considered entanglements are taken to be preserved across consecutive configurations else, the considered entanglements are taken to be different across consecutive configurations. By performing a pairwise comparison between the set of entanglements in earlier configuration with the set of entanglements in later configuration, we can determine the fraction of lost entanglements and the fraction of new entanglements in the later configuration in comparison to the earlier configuration.

While determining the old entanglements and new entanglements between consecutive configurations of the polymer sample, it is important to properly account for the following complication. During the deformation of the polymer sample, the pair of chain segments constituting an entanglement can slide over each other while keeping the entanglement intact. To account for this sliding of chain segments, we permit a range while comparing the monomer positions of the considered entanglements across two consecutive configurations. In order to decide the ideal range to compare the monomer positions of the considered entanglements across two consecutive configurations, we have tested different values of ranges corresponding to 1, 5, 10, 20, 50, 75 and 100 monomers, and find that the qualitative trends for the number of new and lost entanglements between two consecutive configurations across the 10 samples is insensitive to the specific choice of the range (SI Figures S14-S23). For later analysis, we decided to take the range for comparison of the monomer positions of considered entanglements across consecutive configurations to be 50 monomers based on these observations.



\section{Observations from molecular dynamics simulation of the polymer sample: Evolution of craze and void statistics}
\label{obs-md}

\begin{figure}
\includegraphics[width=.8\columnwidth]{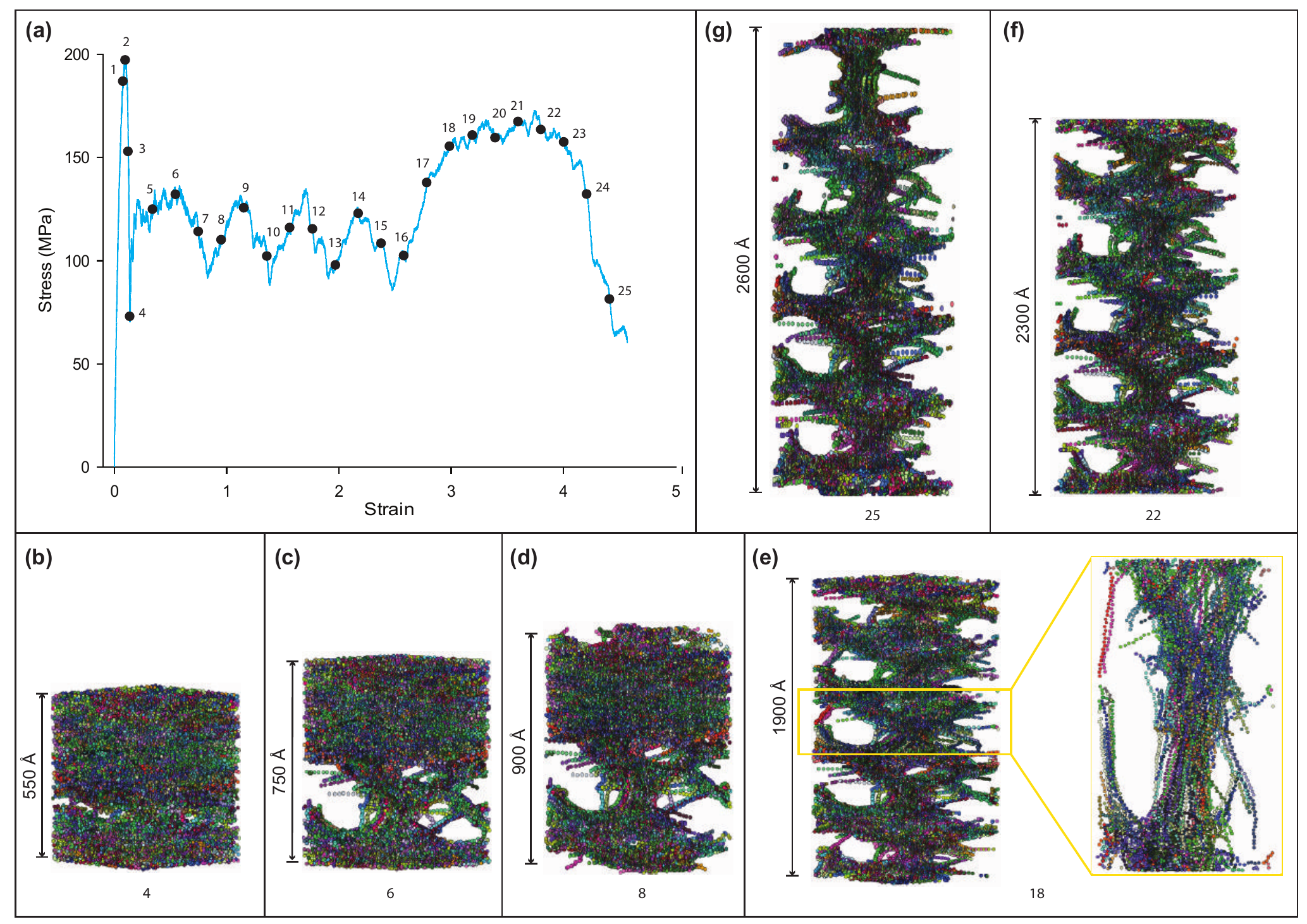}
\caption{The stress-strain curve with configurations showing important stages of craze initiation, growth and failure in the polymer sample. (a) The evolution of the stress in the sample at different strain during polymer deformation. The black points on the stress-strain curve specify the 25 configurations (numbered from 1 to 25) that were selected for the reconstruction and analysis of Van der Waals networks and entanglement networks. The figure also displays configurations of the polymer sample at different stages of deformation process: (b) Initiation of first craze (stage 4), (c) Growth of the first craze (stage 6), (d) Initiation of the second craze (stage 8), (e) Sample after complete crazing (stage 18), (f) Sample at the beginning of the final failure stage (stage 22), and (g) Sample near final failure (stage 25). Note that the configurations of the polymer sample displayed in (b)-(g) have not been drawn to scale.}
\label{stress-strain-curve}
\end{figure}

The results of MD simulations for the 10 different samples successfully captures the crazing phenomena (Figure \ref{stress-strain-curve}) and has high resemblance to the honeycomb craze structure observed in experiments by Duan {\it et al} \cite{Duan1998}. Moreover, we observed that the crazing behaviour and stress-strain characteristics are similar across the MD simulations for the 10 different samples (SI Figure S1). Thus, we report the results from the MD simulation of first sample in the main text, while the results from the MD simulation of remaining 9 samples are included in SI.

Figure \ref{stress-strain-curve} shows the evolution of stress at different strain values for the first sample. The numerical indices at the black points on the stress strain curve are indicative of the indices of the configurations  extracted for network analysis. Figure \ref{stress-strain-curve} also displays select configurations at important phases of polymer deformation such as craze initiation (stage 1 - 4), growth (stage 5 - 15), hardening (stage 16 - 22) and failure (stage 23 -25) in the first sample. At the initial yield point (stage 2) small voids are formed in the first sample which grow and coalesce to form an initial craze at a strain value of 0.14 (Figure \ref{stress-strain-curve}(b)). Thereafter, the craze in the first sample grows by elongation of the fibrils (Figure \ref{stress-strain-curve}(c)). After the growth of this first craze, another craze is found to nucleate upon further deformation of the first sample (Figure \ref{stress-strain-curve}(d)). Figure \ref{stress-strain-curve}(b)-(d) show a cycle of craze initiation, growth and nucleation of subsequent craze in the first sample. In the stress versus strain curve (Figure \ref{stress-strain-curve}(a)), this cycle of craze initiation, growth and nucleation of subsequent crazes can be identified from the drop in the stress after the yield point followed by an increase in the stress which is followed by another drop in the stress value. For example, the stress versus strain curve between strain values 0.547 to 1.158 in Figure \ref{stress-strain-curve}(a) constitutes one such cycle. Moreover, three subsequent cycles are observed between strain values 1.158 to 2.786 in our simulation of the first sample (Figure \ref{stress-strain-curve}(a)). During the beginning of the hardening phase (Figure \ref{stress-strain-curve}(e)), there are four mature crazes and a fifth nascent craze in the first sample (stage 16). As the fifth craze grows in the sample, there is no available room for further craze formation, and the sample with crazes resembles a honeycomb structure (Figure \ref{stress-strain-curve}(e)-(f)). Upon continuous deformation, the polymer sample grows by further stretching of elongated fibrils and exhibits hardening (stage 16 - 20). By this stage of the deformation process, one can assume that the fibers in sample are in fully stretched state, and any further deformation leads to weakening of the entanglement structure due to sliding of chain segments (stage 20 - 22). The beginning of the final failure (stage 23) of the polymer is indicated by the start of the sudden plunge in the stress value (Figure \ref{stress-strain-curve}(f)). The final failure occurs locally in one of the crazed layers as seen in the configuration near final failure in Figure \ref{stress-strain-curve}(g) where craze structure at the top of the sample is far more deformed in comparison to other layers of the sample. Apart from the above observations, the following additional conclusions were made by Venkatesan {\it et al} \cite{Venkatesan2015}. The initial craze formation was typically observed to start in an area of the sample with low entanglement density. As the craze grows, the strong entanglements in the sample tend to cluster into a region near the ends of fibrils. This region with high entanglement density forms strong anchoring layer, and thus, limits the extension of the first craze. Upon continuous deformation, regions adjoining the first craze with low entanglement density also begin to craze. After complete crazing of the entire sample, any further deformation causes disentanglement in the strong anchoring layer leading to final failure of the polymer \cite{Venkatesan2015}.

\begin{figure}
\includegraphics[width=.63\columnwidth]{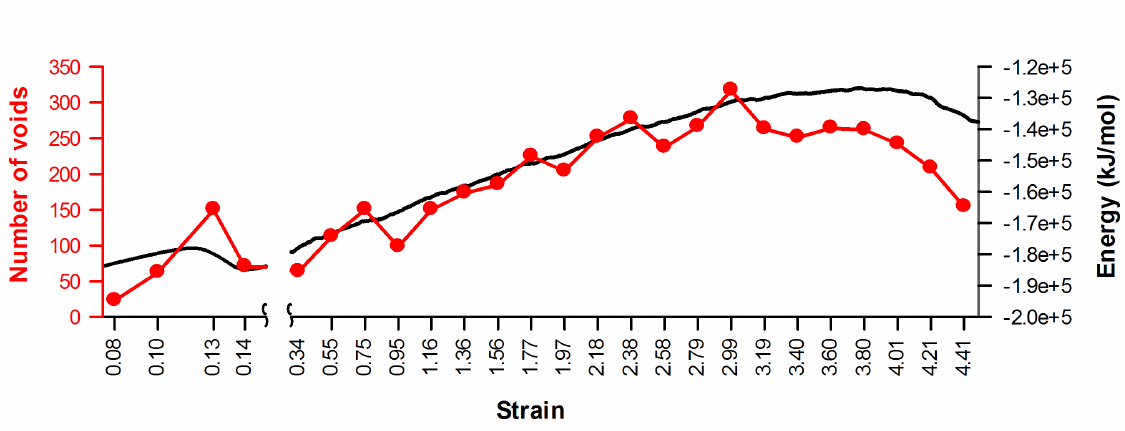}
\caption{The total number of voids and the total intermolecular nonbonded energy from Van der Waals interactions between united atoms at different strains values for the first sample.}
\label{void-energy}
\end{figure}

Void nucleation, growth and coalescence are integral to the crazing phenomena in polymers. Recently, Ichinomiya {\it et al} \cite{Ichinomiya2017} have employed persistent homology to study the role of voids in the initial yielding and softening phases of the crazing phenomena in polymers. Ichinomiya {\it et al} \cite{Ichinomiya2017} concluded that the yielding of polymers is a case of significant structural changes and percolation of voids. In this work, we have also studied the role of voids in the growth and failure phases of crazing phenomena in polymers. Furthermore, we use an exact method detailed in section \ref{void-method} to identify voids and their respective sizes (SI Figures S2-S13). The void statistics showed an increase in the number of voids (Figure \ref{void-energy}; SI Figure S2) after yield point (between the stage 2 and stage 3 in the stress-strain curve of Figure \ref{stress-strain-curve}(a)) followed by coalescence (between the stage 3 and stage 4 in the stress-strain curve of Figure \ref{stress-strain-curve}(a)). The void size distribution shows the appearance of a large void at the end of softening phase, which is indicative of the initial craze formation. During the growth phase, the voids grow in size. In the growth phase as other layers craze in succession, we observe the large voids are not more than two in number which indicates that the large voids span across different layers. As the entire sample crazes and the hardening phase begins, the large voids coalescence and a single large void spans the entire sample. Thus, the void structure percolates the entire sample as the hardening of the sample begins (stage 16), and this feature is universal across the 10 samples. Consistently, a marked jump is observed in the void size distribution when a single void spans the entire sample (SI Figures S2-S13). The number of voids increases almost steadily during the growth phase (stage 5 - 15) while being almost constant during hardening phase (Figure \ref{void-energy}; SI Figure S2). Our observation of increase in void volume with number of voids remaining constant in the hardening phase (stage 16 - 20) indicates growth of voids by expansion. The final failure phase (stage 23 - 25) is characterized by decrease in the number of the small voids, while the largest void grows in size as a result of the relaxation in the rest of the sample while the localized failure region weakens.



\section{Reconstruction of underlying networks from the polymer sample}
\label{reconstruction}

It is well-known that intermolecular forces, namely Van der Waals forces, play a role in the deformation of polymers \cite{Mark1942, Termonia1987}. In addition to Van der Waals forces, the role of entanglements between the polymer chains in the deformation of polymers has been studied. The importance of a stable entanglement network, where the polymer chain length is sufficiently greater than the entanglement length, in the formation of crazes has been discussed by Baljon {\it et al} \cite{Baljon2001}. The role of entanglement network in determining the mode of failure has been discussed by Deblieck {\it et al} \cite{Deblieck2011}. As mentioned earlier, a mesoscale model that incorporates Van der Waals interactions and entanglements between chains can successfully reproduce bulk polymer behaviour \cite{Reddy2008}. Thus it is important to reconstruct the two underlying networks, Van der Waals network and entanglement network, from the polymer sample as they capture the essential interactions governing the polymer behaviour. In this section, we describe the reconstruction of the two underlying networks, Van der Waals network and entanglement network, from the snapshots of the polymer sample at different strain values obtained from the
MD simulation (Figure \ref{netrecon}).

\subsection{Van der Waals network}

The Van der Waals interactions between different united atoms of a dense polymer forms a highly interconnected network. Given the cut off distance of 12 $\AA$ for the non bonded LJ potential in our MD simulation, a typical united atom in a single chain of the polymer can easily interact with at least 50 other neighboring united atoms. Thus, in order to circumvent the computational challenges associated with the analysis of very large and dense networks, we resort to a novel coarse-graining algorithm, where we substitute 10 consecutive united atoms in a single chain by a bead. This substitution was motivated by the measured Kuhn length of 12 $\AA$ for polyethylene melts \cite{Aharoni1983}. Note that Kuhn length measures the intact segment of a polymer chain. Thus, in our Van der Waals network, 1000 united atoms in a polymer chain are coarse-grained into 100 beads, and this effectively reduces the number of Van der Waals interactions. Nevertheless, our coarse-graining algorithm to reconstruct the Van der Waals network preserves the qualitative and quantitative features of the Van der Waals interactions in the polymer \cite{Venkatesan2015}. The strength of Van der Waals interaction between two beads at a distance $r$ from each other in the coarse-grained network is estimated as the summation of the interactions between constituent united atoms. The resulting dataset of energy versus distance for different pairs of beads in the polymer sample is fitted with a Lennard-Jones (LJ) type potential to estimate the Van der Waals interactions (Figure \ref{netrecon}(b)).

In the reconstructed Van der Waals network, coarse-grained beads are taken as nodes. A cut-off distance of 12 $\AA$ on the interbead distance was used to define an edge between two nodes. In the coarse-grained Van der Waals network, the weight of an edge between two nodes depends on the equilibrium distance $r_{eq}$ (at which the LJ potential attains minimum) and the actual distance $r$ between two nodes in the network (Figure \ref{netrecon}(b)). Note that our definition of the edge weights in the coarse-grained Van der Waals network reflects the strength of Van der Waals interaction between two beads, and the edge weight can take a value between 0 and 1. Specifically, if the distance between two nodes in the Van der Waals network exceeds the equilibrium distance, the edge weight is taken as the ratio $\frac{r_{eq}}{r}$ which is less than 1 (Figure \ref{netrecon}(b)). Alternatively, if the distance between two nodes in the Van der Waals network is less than or equal to the equilibrium distance, the edge weight is taken as 1 (Figure \ref{netrecon}(b)). In section \ref{networkanalysis}, we report our results from graph-theoretic analysis of Van der Waals networks reconstructed from the polymer samples (Figures \ref{vdw-network} and \ref{vdw-box}; SI Figures S24-S41).
\subsection{Entanglement network}

Entanglements are topological constraints between different chains of a polymer which restrict their movement. As discussed earlier, entanglements in the class of polymers considered here can be of varied strengths and complexity. Any given chain of the polymer entangles at different points along its length with other chains. The dynamics of the intervening chain segment connecting two entanglements is restricted by the presence of the entanglements. Moreover, the entanglements evolve based on the stretch on the chain segment connecting them. Thus, an entanglement network associated with the polymer emerges from the entanglement points of various chains and intervening chain segments between different entanglements. This entanglement network of connected chain segments influences the deformation characteristics of the polymer. As the polymer deforms, there are changes in the entanglements between various chains and intervening chain segments, and these changes must be reflected in the associated entanglement network.

\begin{figure}
\includegraphics[width=.8\columnwidth]{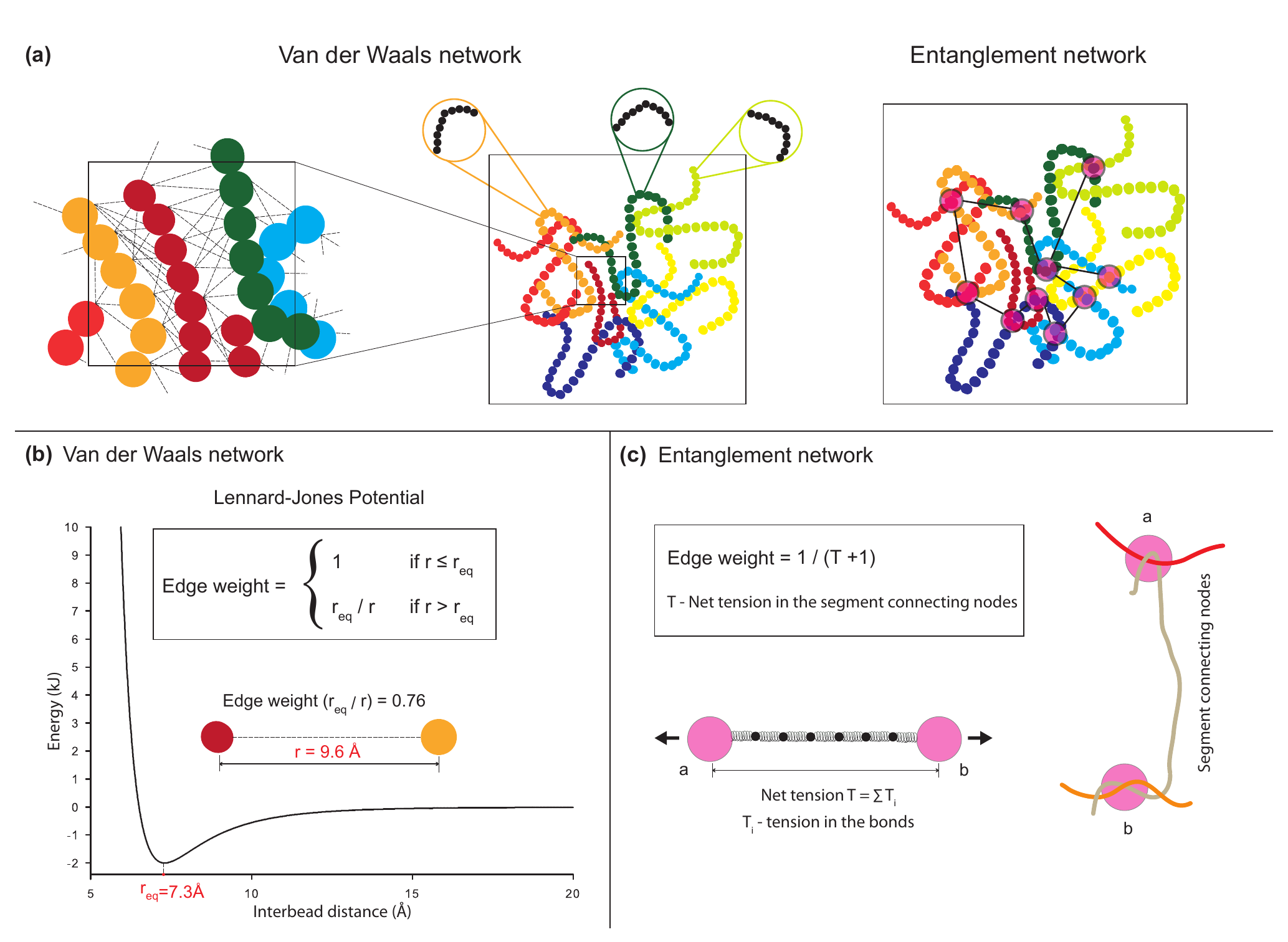}
\caption{Reconstruction of Van der Waals network and entanglement network for the polymer sample. (a) In this schematic figure, a portion of the polymer sample is magnified to display the underlying networks. Every node in the Van der Waals network is obtained by substitution of 10 consecutive monomers on a single chain with a bead. Every node in the entanglement network represents an entanglement between the polymer chains of the sample. (b) Edge weights in the Van der Waals network were determined using actual distance between coarse-grained beads and the equilibrium distance obtained from the associated Lennard-Jones (LJ) potential of the polymer. (c) Edge weights in the entanglement network were determined based on the net tension in the segment connecting two nodes.}
\label{netrecon}
\end{figure}

We reconstruct the entanglement network for the 25 selected configurations of the polymer sample at different strain values as follows. Each entanglement in the polymer sample is represented as a node in the entanglement network (Figure \ref{netrecon}(a)). In the entanglement network, two nodes are connected by an edge if the entanglements are connected by a chain segment (Figure \ref{netrecon}(c)). Note that the chain segments that connect two nodes in the entanglement network can either be in the stretched or coiled configuration (Figure \ref{netrecon}(c)). We remark that the more stretched is the chain segment between two nodes in the entanglement network, the more likely it is for the chain segment to disentangle. Since the bond between united atoms is governed by a harmonic potential (Table \ref{ffield}), the stretch between two entanglements can be quantified by the summation of the tension forces exerted by the bonds constituting the chain segment between connected nodes. Additionally, the angle and dihedral contributions to the tension force between bonded atoms can also be considered. Lastly, we specify the edge weights in the entanglement network as follows. The resultant tension force in the intervening segment connecting two nodes in the entanglement network is determined from the bond, angle and dihedral potentials. Note the tension on the chain segment is positive in the stretched state and negative in the compressed state. We determine the actual bond potential between united atoms constituting the intervening segment by taking the bond constant as unity for simplicity. Similarly, we determine the angle and dihedral potential between united atoms constituting the intervening segment by appropriately scaling the potential constants. This procedure gives the resultant tension force $T$ in the chain segment connecting two nodes in the entanglement network. To ensure that edge weights in the entanglement network reflect the connection strength between two nodes, we specify the edge weight as $\frac{1}{T+1}$ where $T$ is the tension force in the chain segment connecting two nodes (Figure \ref{netrecon}(c)). Note that the connection strength between two nodes in the entanglement network is inversely proportional to the tension force in the intervening segment. Thus, our definition gives edge weight equal to 1 in the entanglement network when there is no tension force in the chain segment connecting two nodes.

We end this section by describing three unusual situations which are encountered during the reconstruction of the entanglement network. Firstly, in situations where the chain segment between two nodes is in a compressed state, and thus, the resultant tension force in the chain segment is negative, we take the edge weight equal to 1 in the entanglement network. Secondly, in some instances, a chain was found to entangle with more than one other chain at the same monomer position, and in such rare cases, the multiple entanglements at the same monomer position were considered as a single node in the entanglement network. Thirdly, it is possible that two nodes in the entanglement network are connected by more than one chain segment, and in such rare situations, the edge weight between the pair of nodes is computed based on the sum of tensions over all segments connecting the two nodes in the entanglement network (Figure \ref{netrecon}(c)).

\begin{figure}
\includegraphics[width=.63\columnwidth]{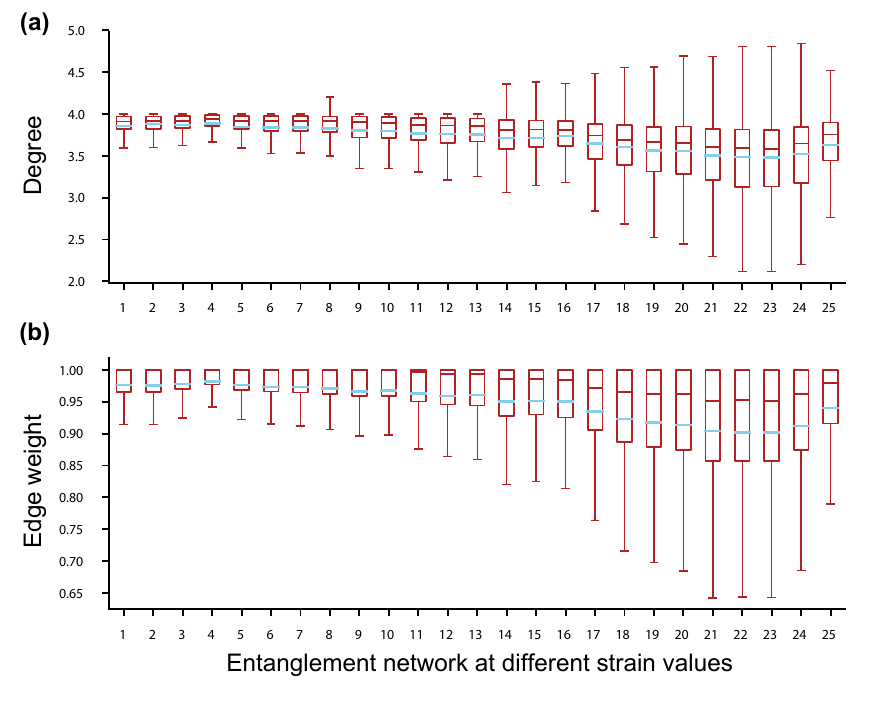}
\caption{Box plot of node degree distribution and edge weight distribution in the underlying entanglement networks of the polymer at different stages of stress-strain curve for the first sample. Note that the label of the 25 entanglement networks considered here is same as those marked in the stress-strain curve of Figure \ref{stress-strain-curve}. The lower end of the box represents the first quartile, brown line inside the box is the median, blue line is the mean and the upper end of the box represents the third quartile of the distribution.}
\label{ent-box}
\end{figure}

In section \ref{networkanalysis}, we report our results from graph-theoretic analysis of entanglement networks reconstructed from the polymer samples (Figures \ref{ent-network} and \ref{ent-box}; SI Figures S42-S79). Note that while computing the edge weights in entanglement networks, we have considered two variations while determining the tension force in the intervening chain segment between two entanglements. Firstly, we have only considered the contribution of bond potential to the tension force, and these results are reported in the main text (Figures \ref{ent-network} and \ref{ent-box}; SI Figures S42-S59). Secondly, we have considered all contributions to the tension force which includes the bond, angle and dihedral potentials, and these results are reported in the SI Figures S60-S79. We remark that our conclusions based on analysis on the entanglement networks hold regardless of the two possible variations to determine the edge weights.


\section{Graph-theoretic analysis of underlying networks associated with polymer}
\label{networkanalysis}

The deformation of the amorphous glassy polymers described earlier and depicted visually in Figure \ref{stress-strain-curve}, suggest reorganization processes involving both short and long length scale fluctuations. In particular, local changes include for example, the breakage and/or formation of local contacts during crazing while large scale changes involve changes in the connectivity of the polymer melt. In order to understand better the different types of fluctuations involved, and to describe the fragmentation process better, we examined the network properties of both the Van der Waals and entanglement networks described earlier in section \ref{reconstruction}. For the underlying networks associated with the 25 polymer configurations extracted from the MD simulation of one initial sample (Figure \ref{stress-strain-curve}), a series of graph-theoretical measures were determined. In the main text, we present the graph-theoretic analysis of the reconstructed networks for MD simulation of one initial sample while the analysis of the simulations of the remaining 9 initial samples is contained in SI.

\begin{figure}
\includegraphics[width=.45\columnwidth]{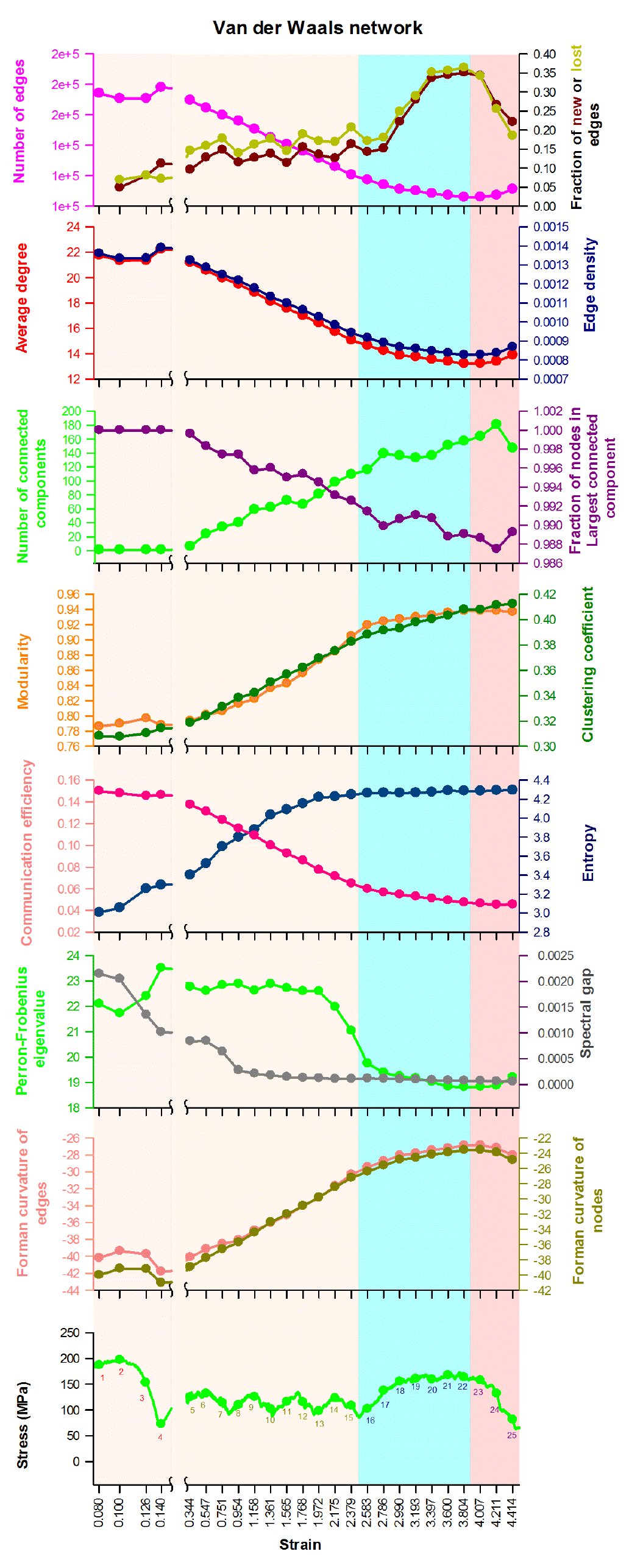}
\caption{Evolution of network measures for the Van der Waals networks at different strain values contrasted against stress-strain curve for the first sample.}
\label{vdw-network}
\end{figure}

\begin{figure}
\includegraphics[width=.45\columnwidth]{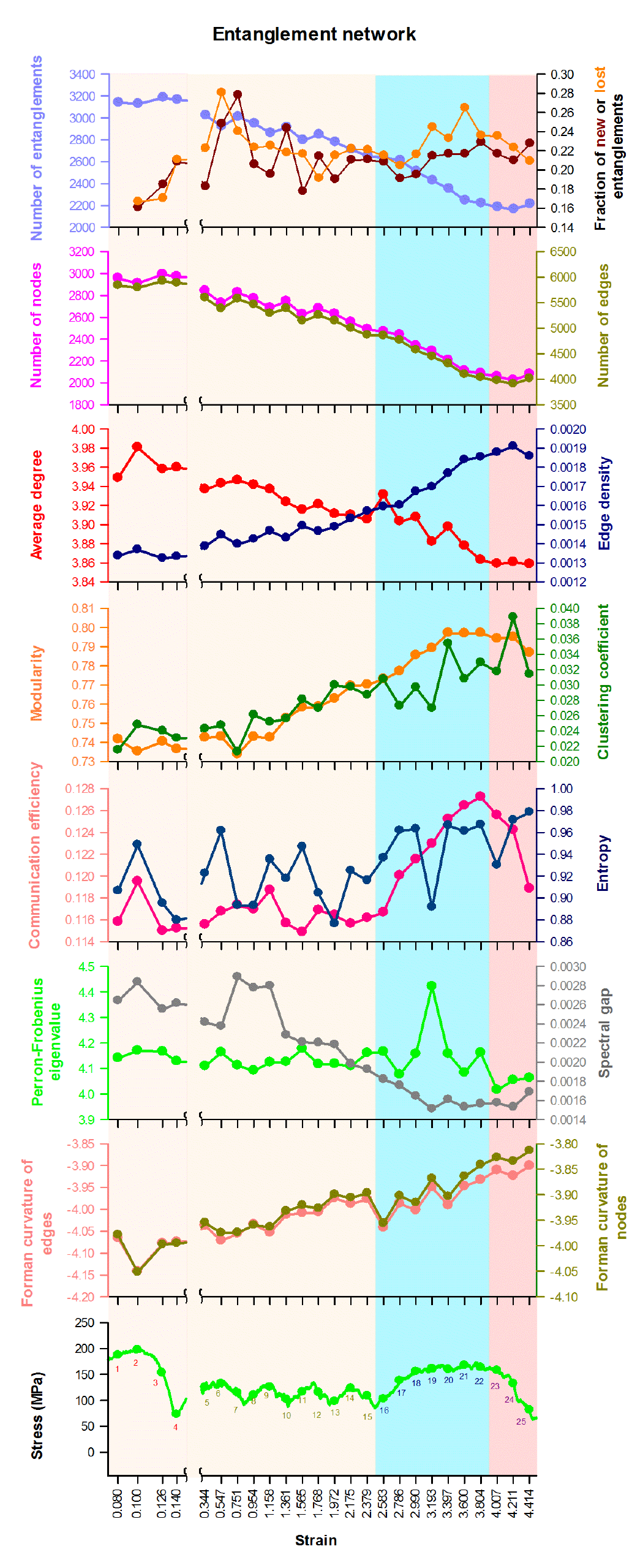}
\caption{Evolution of network measures for the entanglement networks at different strain values constrasted againt stress-strain curve for the first sample.}
\label{ent-network}
\end{figure}

We have used various graph-theoretic measures to characterize the global and local structure of the two underlying networks associated with polymer. In total, 14 graph-theoretic order parameters were examined here. We begin by showing the evolution of the potential energy of the system across the crazing process (Figure \ref{void-energy} and SI Figure S2). At first we show the evolution of the graph properties in the Van der Waals network (Figure \ref{vdw-network}) and then later describe the properties of the Entanglement network (Figure \ref{ent-network}). We see that over the course of the 25 configurations, the potential energy decreases by about 300 kJ/mol per polymer molecule (Figure \ref{void-energy}). The decrease in energy is expected since the number of intermolecular contacts reduces over the course of the crazing. This is reflected in the top 2 panels of Figure \ref{vdw-network} and SI Figures S24-S32, where the number of edges and the average degree of the nodes in the Van der Waals network is shown; in both cases these quantities display an overall decrease. Similarly the fraction of edges formed or lost shows a net loss in edges over the course of crazing (Figure \ref{vdw-network}). However at the end of softening (stage 4) fraction of new edges exceeds the fraction of lost edges indicating relaxation in the polymer sample as strain localisation occurs initiating crazing. The yield (stage 2) is marked by significant fraction of edges  lost. Interestingly, during stages 18 - 22 preceding failure, there is both significant amount of formation of new edges and loss of old edges with no net change in the number of edges in the Van der Waals network. This indicates that there is significant relative movement between the polymer segments preciding failure. In the final phase (stage 23- 25), the decrease in the fraction of new and lost edges in Van der Waals networks suggests that the inter-segmental motion within the polymer has dropped significantly as localized failure occurs.

Another way to probe the network evolution of stage 5 - 15 (Figure \ref{vdw-network}), is through examining the number of connected components, as well as the fraction of nodes in the largest connected component. A connected component is a subgraph of a network, wherein there always exists a path connecting all pairs of nodes in it. Again, we clearly see that these two quantities are anti-correlated in Van der Waals network (Figure \ref{vdw-network}). The number of connected components during the crazing reduces by a factor of approximately 10 between stages 5 and 15. However, despite this, there is only a small change in the percentage of nodes in the largest connected component. This implies that despite the fragmentation during crazing, there is still a large cluster which remains highly connected within the polymer sample.

Panel 4 of Figure \ref{vdw-network} and SI Figures S24-S32 shows the evolution of the clustering coefficient \cite{Watts1998} and the modularity \cite{Newman2006,Blondel2008} of the Van der Waals network during the crazing process. Similar to the quantities discussed above, during the first 4 stages, both the clustering and modularity remain constant before increasing. Interestingly, the  modularity, which probes the community structure of the network increases during crazing which is consistent with the fact that highly modular networks are characterized by high density of intra-module edges and sparse density of inter-module edges. As more voids appear and grow, forming fibrils and tightly packed inter-layers in the polymer sample, the associated restructuring of Van der Waals network is reflected by the trend in modularity. The decrease in the average degree and number of connected components in the Van der Waals network is manifested in the communication efficiency \cite{Latora2001} shown in panel 5 of Figure \ref{vdw-network} and SI Figures S24-S32. Essentially this measure quantifies how easy it is for information to propagate in the network as probed by the inverse of the shortest path length between all pairs of nodes. Note that in the Van der Waals network the edges are weighted by the strength of Van der Waals interaction between two beads and hence a decrease in the communication efficiency reflects the extent of fragmentation of the Van der Waals network.

\begin{figure}
\includegraphics[width=.63\columnwidth]{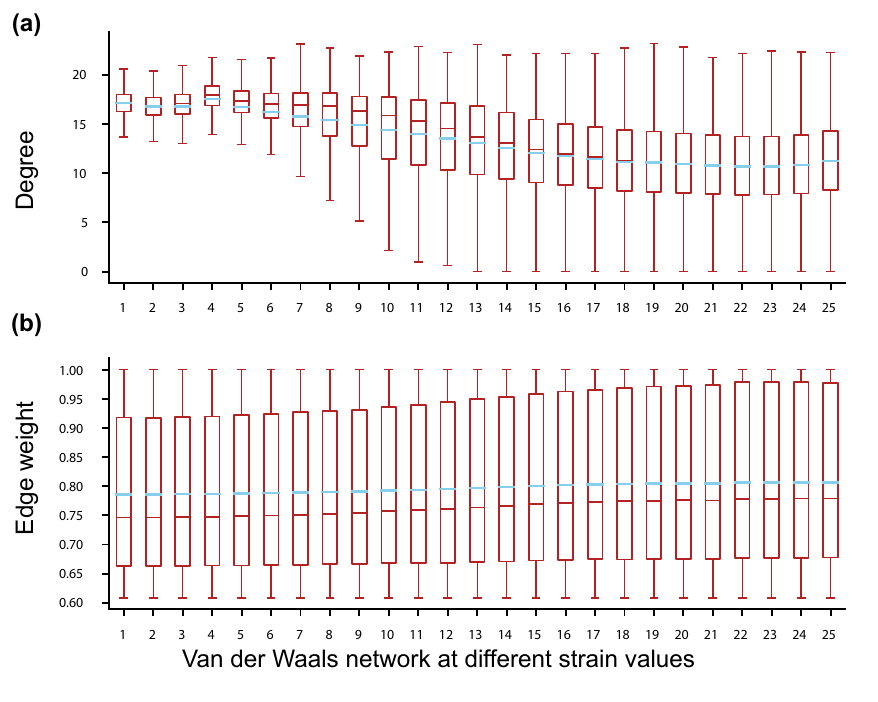}
\caption{Box plot of node degree distribution and edge weight distribution in the underlying Van der Waals networks of the polymer at different stages of stress-strain curve for the first sample. Note that the label of the 25 Van der Waals networks considered here is same as those marked in the stress-strain curve of Figure \ref{stress-strain-curve}. The lower end of the box represents the first quartile, brown line inside the box is the median, blue line is the mean and the upper end of the box represents the third quartile of the distribution.}
\label{vdw-box}
\end{figure}

In all the graph-theoretic measures elucidated in the preceding paragraphs, during the first 4 stages all the parameters remain constant for Van der Waals networks although it is clear from the stress-strain curve that there are big changes in the initial stage 1 to 4 (Figure \ref{vdw-network}). We remark that yielding is mainly influenced by local weaker regions within the polymer sample, and this may explain the absence of any striking change in global network measures such as number of connected components and communication efficiency in the craze initiation phase. We thus turned to some other measures like the network entropy as defined by Sole and Valverde \cite{Sole2004} as well as spectral properties, namely, Perron-Frobenius eigenvalue of the adjacency matrix and the spectral gap of the normalized Laplacian matrix. Interestingly, for all these three quantities we observe changes during stages 1 through 4 where the entropy increases and the spectral gap decreases in the craze initiation phase. This implies that the initial uniform network has transitioned to a more heterogeneous partitioned network. The evolution of the entropy is also reflected in the box plot on degree distribution where the range of degree distribution increases as first craze grows (Figure \ref{vdw-box} and SI Figures S33-S41). Subsequently, the first and third quartile of degree distributions increase in width as more of the sample crazes signifying a larger variation in the connectivity of nodes within the network in comparison to the mean value. As the final craze begins to nucleate the entropy of Van der Waals networks reaches a plateau, indicating that the degree distribution has saturated (Figure \ref{vdw-network}). This is reflected in the box plot (Figure \ref{vdw-box} and SI Figures S33-S41), where the difference between the mean and median for the degree distribution vanishes towards final failure stage.

Finally, we also investigated the Forman curvature recently proposed \cite{Sreejith2016,Sreejith2017} to examine the \emph{geometry} of the networks. Here we observe that this quantity increases with crazing in Van der Waals network consistent with our earlier observation \cite{Sreejith2016,Sreejith2017} that the change in this quantity reflects a decrease in the large-scale connectivity of the network (Figure \ref{vdw-network} and SI Figures S24-S32). The Forman curvature begins to plateau at the end of growth phase (stage 15) and the beginning of the hardening phase (stage 16). In fact, the Forman curvature of all the 10 samples exhibit the same range of values between -30 to -28 as hardening begins. Another interesting observation is that the trend in average Forman curvature for both the nodes and edges reflect closely the trend in total Van der Waals energy of the polymer sample estimated from simulation. Forman curvature also has an inverse correlation with the average degree, number of edges and the edge density in Van der Waals networks (Figure \ref{vdw-network}). The end of the growth phase which is also the beginning of the hardening phase is clearly signified by these parameters reaching similar values across the 10 samples irrespective of the applied strain.

The preceding analysis focuses on the graph-theoretical measures from the Van der Waals networks. In order to understand if the Entanglement networks behave in a similar or different manner to the Van der Waals networks, we show in Figure \ref{ent-network} and SI Figures S42-S50 the same measures for the former. An important point to note is that in the entanglement networks, the network evolution involves more collective or larger scale correlations in the network as described earlier. As seen in Figure \ref{ent-network} and SI Figures S42-S50, the number of nodes and average degree in the entanglement network behaves in a very similar manner to the Van der Waals networks. This is also true for measures such as the modularity, Forman curvature and the clustering coefficient. The major differences we see are in the communication efficiency, entropy and spectral properties (Figure \ref{ent-network} and SI Figures S42-S50). For the entanglement network, the communication efficiency remains more or less constant till stage 16 before increasing (Figure \ref{ent-network}). It is worth pointing out that in these networks the paths are weighted by the tension in the polymer chain as defined earlier. This suggests that there are entanglements or connections in the network with strong tensile forces up to much higher strain than that revealed by the Van der Waals network. The box plots for the edge weight distributions (Figure \ref{ent-box} and SI Figures S51-S59) in entanglement networks clearly show a marked drop in the stage 16 - 23 indicating a increase in the tension in the intervening segments between entanglements. As the final phase (stage 23 - 25) is signified by localized failure in one of the layers, the rest of the sample relaxes. This relaxation of the entanglement network leads to an increase in the mean edge weight (Figure \ref{ent-box} and SI Figures S51-S59).

Another interesting feature of the entanglement networks is the behavior of the network entropy. Recall that the entropy of a network is defined as an average measure of network's heterogeneity which probes the diversity of the link distribution \cite{Sole2004}. In the case of entanglement networks, the entropy undergoes large fluctuations during crazing but does not change by as significant amount as it did in the Van der Waals network (Figures \ref{vdw-network} and \ref{ent-network}). Similarly, the Perron-Frobenius eigenvalue remains relatively constant until stage 17-20 where there is a large fluctuation just like in the Van der Waals network and the spectral gap drops overall during the crazing process (Figure \ref{ent-network} and SI Figures S42-S50). The spectral gap, in essence, quantifies the extent of bottlenecks in graphs and the associated eigenvector is utilized in graph partitioning algorithms. Since there is an overall decrease in the spectral gap of the underlying networks during crazing, this suggests that the underlying networks become more fragmented in the course of crazing.


\section{Summary and Future Outlook}
\label{summary}

In this work, we simulate an amorphous polymer sample using molecular dynamics simulations to study its crazing behaviour. The resulting stress strain behaviour and the extracted molecular configurations of the sample at different stages of crazing are used to probe the physical phenomenon. The development of the void structure which is the prominent plastic deformation characteristic is investigated throughout the entire crazing process.

The trend in void evolution, growth and coalescence was found to be reflected by the trend in the fraction of new edges or lost edges in the Van der Waals network. Our results on void evolution during the initial yielding phase was in accordance to the recent findings of Ichinomiya {\it et al} \cite{Ichinomiya2017} who used a method based on persistent homology to identify voids in the polymer sample. However, Ichinomiya {\it et al} \cite{Ichinomiya2017} have investigated the evolution of void structure only in the initial craze formation stage while we have investigated throughout the entire crazing process. Furthermore, we use an exact method to estimate voids in the polymer sample that circumvents the difficulty in estimating the three dimensional structural size of voids using the persistent homology based method \cite{Ichinomiya2017}.

Given the importance of energetic interactions and topological constraints between the polymer chains in determining the deformation characteristics, we use network theory as a tool to analyze and interpret the results from MD simulation of the crazing phenomena in a glassy amorphous polymer. We have investigated the dynamics of the crazing phenomena by reconstructing two underlying networks associated with a polymer sample, Van der Waals network and entanglement network, for selected configurations across different stages of deformation obtained from MD simulations. Our results from the analysis of underlying networks associated with a polymer attest that network measures successfully capture the underlying physics governing the different stages of crazing in polymers.

The Van der Waals network plays a prominent role in the craze initiation (stage 1 - 4) and growth (stage 5 - 15) during polymer deformation (Figure \ref{vdw-network}). Yielding of the polymer sample is characterized by significant structural changes in the Van der Waals network (Figure \ref{vdw-network}). During growth stage, the Van der Waals network evolves into a modular network (Figure \ref{vdw-network}). Beginning of the hardening (stage 16) is captured by marked changes in a number of graph-theoretic measures in Van der Waals network, and this change is irrespective of the strain value at which hardening starts (Figure \ref{vdw-network}). This is a significant observation, since during the development of continuum models, the beginning of the hardening stage can be revealed using the characteristics of the Van der Waals network. The fact that the energetics of a material (represented by the Van der Waals network) governs the important transitions rather than a fixed stress or strain value is elaborated from the experimental studies \cite{Bruller1991} on many polymer deformation characteristics including crazing phenomena. The Perron-Frobenius eigenvalue for the Van der Waals network shows specific signatures in the softening phase (stage 2 - 4) and at the transition from the craze growth to the hardening phase (stage 15 to stage 16) (Figure \ref{vdw-network}). The entanglement network evolves during the growth stage (stage 5 - 15), and plays a significant role in the hardening and subsequent failure (stage 16 - 23) (Figure \ref{ent-network}). The role of the entanglement network in the hardening and subsequent failure stages is inferred from communication efficiency and loss of entanglements. The Forman curvature was found to exhibit an inverse correlation with the average degree of nodes in both Van der Waals networks and entanglement networks, and thus, the Forman curvature is an important order parameter to characterize polymeric networks.

We conclude by suggesting some future directions. In the present study, we have analyzed the Van der Waals network and entanglement network for the complete polymer sample at different stages of deformation. This work has successfully used network theory to analyze polymer samples and draw significant conclusions on crazing phenomena. We have identified specific network measures which show marked variation at important stages of the crazing process in the Van der Waals network and entanglement network. Apart from yielding and hardening, localized deformations govern the evolution of crazing phenomena. Since several network measures such as degree and clustering coefficient are associated to every node, the evolution of such measures in localized regions will offer more insight into the crazing phenomena. Also, it will be interesting to investigate crazing in different polymer systems using our framework of Van der Waals network and entanglement network, and this will lead to a better understanding of the significance of the various network measures. Though our present study has focussed on crazing, other deformation processes undergone by polymeric materials are equally important. Under the effect of a constant load, polymers undergo creep failure over a extended period of time. Polymer components subjected to periodic cycles of loading and unloading, eventually break. Shear deformation is another prominent failure mechanism where the polymer fails under uniaxial loading. Hence, depending on the loading conditions, the nature of deformation mechanism exhibited by the polymer also varies. The nature of the response of the Van der Waals network and entanglement network to the different deformation processes like shear deformation, creep loading and cyclic loading can also be studied using our framework. Any specific signatures in the network-based order parameters observed across different deformation processes and polymer systems will be of considerable research interest.

In future, it may be possible to use the insights obtained from network analysis of polymer samples to develop mesoscopic models with simplified representations of the polymer. Previously \cite{Termonia1987}, a mesoscale model based on the Van der Waals interactions and the entanglement between chains has been proposed, and this model was extended \cite{Reddy2008} to study the response of the finite element model to a constant volume deformation. Qualitative predictions by Reddy {\it et al} \cite{Reddy2008} were representative of typical polymer behaviour. The governing parameters like activation energy, activation volume, entanglement strand length used in the above mesoscale model, regulate the breaking of lumped Van der Waals interactions between chains and the stretch in the entanglement segments. Using network analysis of the MD simulations, we may be able to develop more realistic mesoscale models for specific polymers of interest. For example, an improved mesoscale model for an amorphous polymer may be designed such that it reflects the edge weight distribution observed in our analysis of Van der Waals network and entanglement network. In conclusion, our results from network analysis of crazing in glassy amorphous polymers will stimulate future applications of network science in understanding the mechanical properties of polymers.

\section*{Acknowledgments}
The authors thank J\"urgen Jost for discussion and M. Karthikeyan for help with figures. SV would like to thank Sumit Basu for introducing her to molecular dynamics simulations to study crazing in polymers. AS would like to acknowledge financial support from Department of Science and Technology (DST) India [Ramanujan fellowship (SB/S2/RJN-006/2014); Start-up grant (YSS/2015/000060)] and Max Planck Society Germany [Max Planck Partner Group on Mathematical Biology].

%


\end{document}